\documentclass[useAMS,usenatbib,usegraphicx]{mn2e}

\title[Radio recombination lines from the largest bound atoms in space]{Radio
recombination lines from the largest bound atoms in space}
\author[S.V. Stepkin et al.]{S. V. Stepkin$^{1}$\thanks{E-mail:
s.v.stepkin@ira.kharkov.ua },  A. A. Konovalenko$^{1}$,
\newauthor N. G. Kantharia$^2$ and N. Udaya Shankar$^3$ \\
$^1$ Institute of Radio Astronomy, NASU, 4 Chevonopraporna St, Kharkov 61002, Ukraine \\
$^2$  National Centre for Radio Astrophysics, TIFR, PB 3 Ganeshkhind, Pune 411007, India \\
$^{3}$Raman Research Institute, C. V. Raman Av., Sadashivanagar, Bangalore 560080, India }

\begin{document}

\date{}

\pagerange{\pageref{firstpage}--\pageref{lastpage}} \pubyear{}

\maketitle

\label{firstpage}

\begin{abstract}
In this paper, we report the detection of a series of radio
recombination lines (RRLs) in absorption near 26 MHz arising from
the largest bound carbon atoms detected in space. These atoms,
which are more than a million times larger than the ground state
atoms are undergoing delta transitions ($n\sim1009$, $\Delta n=4$)
in the cool tenuous medium located in the Perseus arm in front of
the supernova remnant, Cassiopeia A. Theoretical estimates had
shown that atoms which recombined in tenuous media are stable up
to quantum levels $n\sim1500$.  Our data indicates that we have
detected radiation from atoms in states very close to this
theoretical limit. We also report high signal-to-noise detections
of $\alpha$, $\beta$ and $\gamma$ transitions in carbon atoms
arising in the same clouds.  In these data, we find that the
increase in line widths with quantum number ($\sim \propto n^5$)
due to pressure and radiation broadening of lines is much gentler
than expected from existing models which assume a power law
background radiation field.  This discrepancy had also been noted
earlier. The model line widths had been overestimated since the
turnover in radiation field of Cassiopeia A at low frequencies had
been ignored. In this paper, we show that, once the spectral
turnover is included in the modeling, the slower increase in line
width with quantum number is naturally explained.

\end{abstract}


\section{Introduction}

Over the forty years after their first detection, RRLs have become
an important probe for interstellar plasma investigations
\citep{b5}. The detection \citep{b7,b1,b8} of the low frequency
RRLs of carbon in absorption towards the strong supernova remnant
Cassiopeia A (Cas A)  in 1980 opened new ways to study the tenuous
and cool interstellar medium (ISM). Theoretically, bound atoms
with $n\sim 1500$ \citep{b13} are expected to exist in tenuous
media. Alkali and alkaline earth metals like potassium and barium
have been excited to high quantum levels (up to $n\sim1,100)$ in
the laboratory by using tunable dye lasers and reducing the Stark
effect by minimizing local electric field \citep{b4}. Naturally,
the question arose - could such large atoms be observed with radio
astronomical methods?  Since atoms in such high quantum states
trace entirely different ionized regions compared to HII regions,
they are also important for studying physical processes in the
ISM.  RRL observations till date \citep{b6,b9,b10,b14} cover the
range from n=166 to n=868 towards Cas A with the highest bound
state of $n=868$ near 20 MHz observed by \citep{b9} using the
UTR-2 telescope near Kharkov in Ukraine. At such low radio
frequencies, terrestrial radio frequency interference has to be
excised and data collected over many days to obtain a high signal
to noise detection of any feature since the system temperature is
fairly high due to the non-thermal nature of the background
radiation field. In addition, the lines broaden due to non-LTE
(local thermodynamical equilibrium) effects and the frequency
separation between features ($ 1/n^4$) reduces at low frequencies
making the observations difficult.

In spite of all these difficulties, low frequency RRLs are being
observed in many Galactic regions. In this paper, we report the
detection of the largest bound atom in space with transitions near
$n\sim 1009$ near 26 MHz towards Cas~A.

While lines at frequencies above 110 MHz are observed in emission,
which implies the importance of stimulated emission \citep{b10},
the lower frequency features appear in absorption against the
radiation field of Cas~A.  Since RRLs have been detected at a
range of frequencies from the clouds in the Perseus arm in front
of Cas~A, these have been extensively used to model and understand
the physical conditions in the line forming clouds. Since Cas~A
dominates the system temperature at all these frequencies, it has
been easy to use data from widely different telescopes.  It is now
widely believed that these low frequency RRLs arise in clouds
associated with the HI component of the interstellar medium with
temperatures of $50-75$ K \citep{b6,b11,b14}. However, most of the
models which well explain the variation in the line optical depth
with quantum numbers, overestimate the line widths at low
frequencies.

In this paper, we explain the discrepancy between the observations
and the existing models used to derive the line widths.  In
section \ref{sec2} we elaborate on the observations, data analysis
and present the spectra. In section \ref{sec3}, we discuss our
results and the width discrepancy and in section \ref{sec4} we
present the conclusions of our study.

\section[]{Observations}
\label{sec2}

The RRL observations towards Cas~A were conducted using the
biggest decametre wavelength radio telescope UTR-2 situated in
Ukraine \citep{b2}. UTR-2 is a T-shaped antenna consisting of 2040
fat dipoles.  It has an effective collecting area of about 100,000
m$^2$ which results in a signal from Cas A which is $\sim10$ dB
higher than the Galactic background at 26 MHz.  The telescope has
an electronic steering capability which can track upto
$\pm70\degr$ about the zenith.

The observations were carried out between 2000 and 2004 for
approximately 100 days.  A 4096-channel autocorrelometer operating
with a bandwidth of about 1.2 MHz served as the backend.

\begin{figure}
\includegraphics[width=84mm]{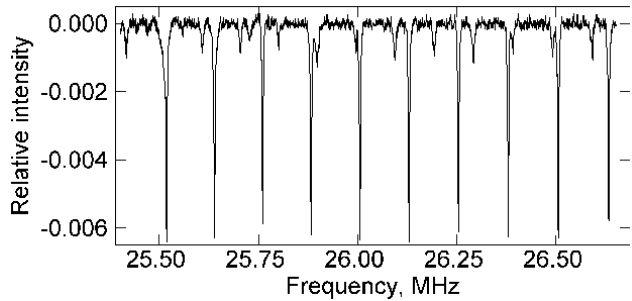}
\caption{Spectrum obtained in the direction of Cas A with UTR-2.
Relative intensity means ratio $T_{L}/T_{c}$. $\alpha$, $\beta$
and $\gamma$ lines are visible.} \label{fig1}
\end{figure}

\begin{figure}
\includegraphics[width=84mm]{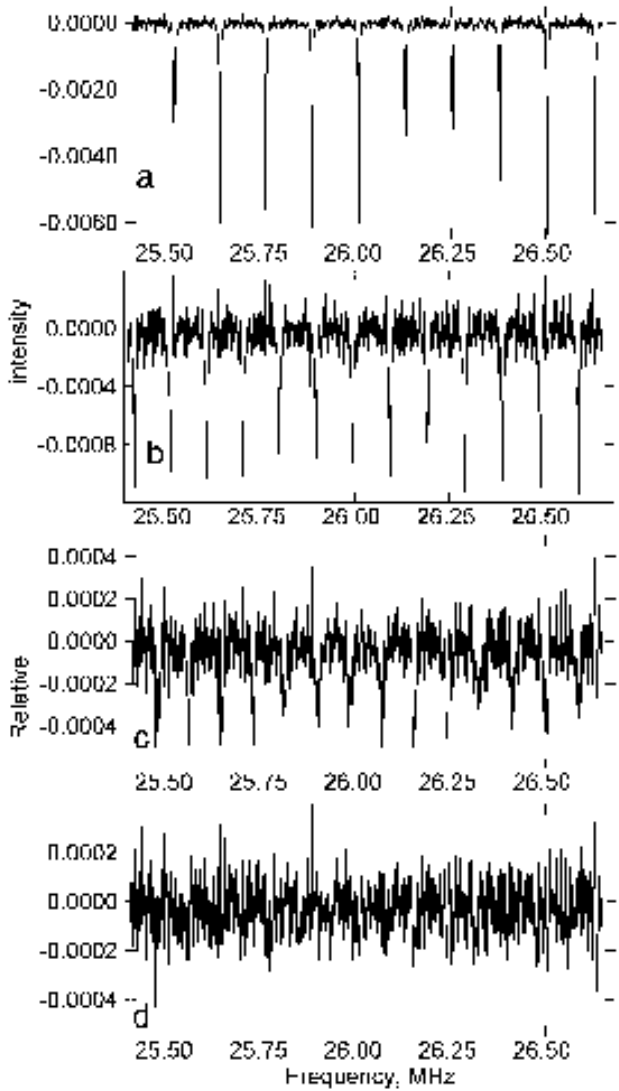}
\caption{ The series of $\alpha$, $\beta$, $\gamma$, and $\delta$
radio recombination lines observed around 26 MHz in the direction
of Cas A. Panel (a) shows $\alpha$ series C627.. C636, panel (b)
shows $\beta$ series C790 .. C802, panel (c) shows $\gamma$ series
C904 .. C917, and the panel (d) shows $\delta$ series C994 ..
C1009.} \label{fig2}
\end{figure}

In our case of very high atom quantum states the condition of
$n>>\Delta n$ prevails and hence adjacent features were considered
to be equivalent. Thus, we could fold individual transitions and
improve measurement sensitivity considerably. Moreover, such an
approach provides more reliable line parameters because different
transitions were observed simultaneously with the same antenna and
ambient medium conditions. Fig.\ref{fig1} shows the resulting
spectrum measured in the direction of Cas A in the frequency band
around 26 MHz with UTR-2. The integration time is 504 h. In the
result obtained from each individual session, four-minute frames
were cleared from interferences by subtracting from the measured
autocorrelation function,  cosine functions whose parameters
correspond to hindering signals revealed in the power spectrum.
Most often their numbers were between 4-8. Frames with wide-band
interferences (which were rare) were excluded from analysis. Base
line correction was carried out using splines of the third order.
Unblended $\alpha$ and $\beta$ features were used to obtain the
initial models of the corresponding line profiles. After removing
of $\alpha$ and $\beta$ lines from the initial spectrum and new
base line correction we obtained the $\gamma$ lines and were able
to build the corresponding model. After this, all obtained models
($\alpha$, $\beta$, and $\gamma$ ones) were again subtracted from
the initial spectrum and base line correction repeated. Models of
the features were built by fitting  Voigt profiles to folded
transitions using Levenberg - Marquardt method. The $\delta$ line
model was obtained similarly to the $\gamma$ one.   For $\alpha$
and $\beta$ profile fitting the two component model with central
radial velocities $V_{lsr}$ (comparatively to the local standard
of rest) of  -1.6 and -47.2 km$s^{-1}$ was used. The $\gamma$, and
$\delta$ features were fitted with one component profile (at -47.2
km$s^{-1}$). The values of central radial velocities as well as
the value of Doppler width $\Delta V_{D}$ of 15.8 km$s^{-1}$ were
obtained by two  Voigt profiles fitting to the folded $\alpha$
line and were fixed when modeling the other features. The
procedure described above was carried out several times until the
fitted parameters were stabilized. The whole numbers of these
steps came to six. The Voigt models were used as the radiative and
pressure mechanisms lead to Lorentzian profile and the Doppler
thermal broadening to a Gaussian. The resulting line shape is
described by the Voigt function which is the convolution of the
above mentioned curves. Fig.\ref{fig2} reveals series of $\alpha$,
$\beta$, $\gamma$, and $\delta$ features. They were obtained as a
consequent of subtracting the corresponding model profiles of RRLs
from the original spectrum (see Fig. \ref{fig1}).

\begin{figure}
\includegraphics[width=84mm]{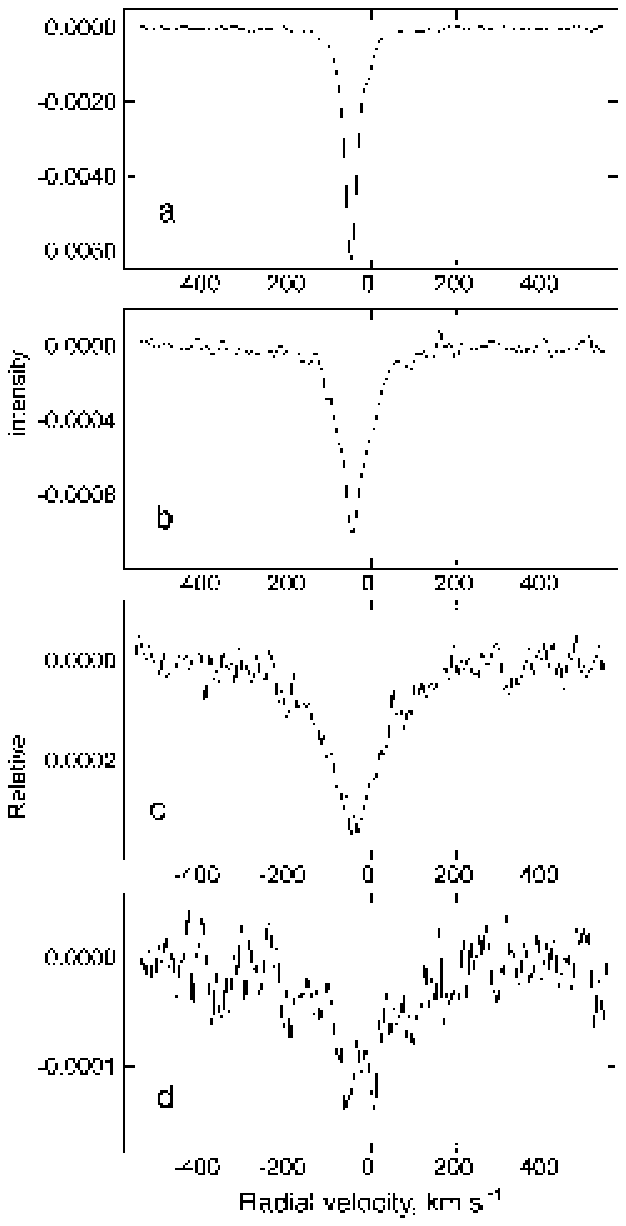}
\caption{The averaged $\alpha$, $\beta$, $\gamma$ and $\delta$
transitions.  Panel (a) shows folded $\alpha$ transitions C627..
C636, panel (b) shows folded $\beta$ transitions C790 .. C802, panel
(c) shows  folded $\gamma$ transitions C904 .. C917, and the panel (d)
shows folded $\delta$ transitions C994 .. C1009.}
\label{fig3}
\end{figure}

The individual transitions in Fig. \ref{fig2} are folded to obtain
the averaged $\alpha$, $\beta$, $\gamma$ and $\delta$ RRLs of
carbon shown in Fig.\ref{fig3}

The feature near $-45$ km s$^{-1}$ is due to gas in the Perseus
arm. In the $\alpha$-line spectrum the spectral feature near 0
km/s corresponds to the medium in the Orion arm. It is blended
with the line arising in the Perseus arm and is not the topic of
consideration in this paper. The effective integration time for
$\alpha$ transition is 5040 h and radial velocity resolution is
9.2 $km s^{-1}$.

The Voigt fitting parameters of lines arising in the Perseus and
Orion arms are presented in Table 1 (for Orion arm features
$\Delta V_{D}=4.4\pm1.1 \  km s^{-1}$).

\begin{table*}
\centering
\begin{minipage}{150mm}
\caption{Voigt fit to the observed line profiles in Perseus arm }
\begin{tabular}{c|c|c|c|c|c}
\\
\hline Galactic arm & Lines & $T_{L}$/$T_{c}$ & $\Delta V_{L}$ &
$\Delta V_{t}$ & I\\ \hline Perseus & $\alpha$(627-636) &
$-0.0061\pm0.00003$ & $29.0\pm2.9$ & $36.2\pm3.6$ & $-28.2\pm2.8$
\\
 & $\beta$(790-802) & $-0.00088\pm0.00003$ & $64.1\pm10.7 $ &
$68.0\pm11.5$ & $-8.0\pm1.5$ \\ & $\gamma$(904-917) &
$-0.00032\pm0.00003$ & $145.1\pm32.6$ & $ 147.0\pm33.1$ &
$-6.3\pm1.4$\\ & $\delta$(994-1009) & $-0.000086 \pm0.00003$ &
$221.5\pm74.6$ & $ 222.8\pm75.1$ & $-2.9\pm1.0$
\\
Orion  & $\alpha$(627-636) & $-0.00044\pm0.00003$ & $16.2\pm3.5$ &
$17.37\pm3.6$ & $-0.9\pm0.8$
\\ & $\beta$(790-802) & $-0.00014\pm0.00003$ & $48.2\pm8.4 $ &
$48.7\pm8.5$ & $-0.8\pm0.8$ \\ \hline
\end{tabular}

\medskip
Symbols used: $T_{L}/T_{c}$ = ratio of line temperature to that of
background (relative intensity);  $\Delta V_{L}$= Lorentzian line
width (km $s^{-1}$); $\Delta V_{t}$= total profile line width (km
$s^{-1}$); I = integral intensity (Hz).
\end{minipage}
\end{table*}

\section{Discussion}
\label{sec3}

Previously, the physical models of the medium in the direction of
Cas A where low frequency carbon spectral lines arise have been
built based on the analysis of the multi frequency data of
$\alpha$ transitions because corresponding features are the most
intensive and easy to observe \citep{b11,b6}. The model which best
fits the data has an electron temperature $T_{e}$=75 K, electron
concentration $n_e=0.02$ cm$^{-3}$ and a background radiation
field $T_{bg}=3200$ K at 100 MHz (spectral index $= -2.6$).
Dielectronic-like recombination \citep{b15} and the boundary
condition that the departure coefficient $b_{n}'\rightarrow 1$ as
$n \rightarrow \infty$ \citep{b12} was used for the modeling.

In order to obtain the maximum excitation level of an atom which
can be studied with the radio astronomical methods, we have to
analyse the problem of low frequency RRLs broadening. This is also
important for explanation of line parameter behaviour. Pressure
and radiative mechanisms are responsible for the effect along with
the Doppler thermal broadening. The Lorentzian line width has two
components
\begin{equation}
  \Delta V_{L}=\Delta V_{P}+\Delta V_{R},
\end{equation}
where $\Delta V_{P}$ and $\Delta V_{R}$ are the pressure and the
radiation broadening correspondingly. Towards Cas A radiation
broadening is caused by influence of the Galactic background with
temperature $T_{bg}$ and non-thermal radiation of Cas A with
$W_{\nu}T_{CasA}$ ($W_{\nu}$ is dilution factor) and $T_{CasA}$ is
the temperature of Cas A radiation, i.e
\begin{equation}
  \Delta V_{R}=\Delta V_{Rbg}+\Delta V_{RCasA}.
\end{equation}
The theoretical behaviours of these phenomena have already been
studied \citep{b13}. The radiation broadening is determined by
\begin{equation}
  \Delta V_{R} \approx 2 c/\pi \nu\sum_{m=1}^{n-1} I_{\nu}B_{n,m},
  km s^{-1}
\end{equation}
where $I_{\nu}$ is intensity of background radiation and $B_{n,m}$
is Einstein coefficient of induced emission, $c$ is light velocity
in km s$^{-1}$ and $\nu$ is the observation frequency in Hz
\citep{b5,b13}.

All models till date have used a power law of index $-2.6$ to
derive the background temperature. In this case
\begin{equation}
  \Delta V_{R}=8*10^{-20}T_{t}n^{5.8}c/\nu,  km s^{-1}
\end{equation}
where $T_{t}=T_{bg}+T_{CasA}W_{\nu}$ and $T_{bg}$ is the radiation
temperature of the background emission at 100 MHz (spectral index
$= -2.6$) \citep{b13}. When calculating the Lorentzian line width
\citep{b6} the total temperature leading to the radiation
broadening was taken as $T_{t}=3200 K$ at 100 MHz ($T_{bg}=800K$)
with spectral index -2.6. This, we believe, is not correct for
quantum numbers $> 650$ or so. Such models give a large
discrepancy between calculated values and experimental data. We
resolve the problem as follows. Cas A has a known turnover point
in its spectrum at frequencies less than 20 MHz. The flux density
drops \citep{b2} from 58,000 Jy at 20 MHz to 26,000 Jy at 10 MHz
and hence instead of increasing with a decrease in frequency, the
flux density decreases. Earlier models had not taken into account
this turnover in the spectrum of CasA and hence overestimated the
line widths at lower frequencies. Here we also have taken
$T_{bg}=800K$ when calculating the contribution of Galactic
background to the total line width. In order to determine the
contribution of Cas A at the low frequencies under study here we
have used the experimental data for frequency range of 10 - 20 MHz
\citep{b2} and extrapolated them up to 5 MHz. The dependence of
flux density and temperature (taking into account the dilution
factor) can be approximated with good accuracy in this range by
\begin{equation}
  S_{CasA} \propto \nu^{1.0},\qquad
  T_{CasA}W_{\nu} \propto \nu^{-1.0}.
\end{equation}
To evaluate the $W_{\nu}T_{CasA}$ we have used the measurement
result for n around 630 ($\nu$ around 26 MHz). Then we calculated
numerically the values of $\Delta\nu_{RCasA}$ for other
frequencies according to (3). In Fig.~4 this result is graphically
presented. The dependence of line width against quantum level
using our algorithm (thick line) and earlier models (thin line)
along with the experimental data are presented. The parameters are
taken as the following: $T_{e}$=75 K, $n_e=0.02$ cm$^{-3}$, the
temperature of the Galactic non-thermal background radiation
T$_{bg}$= 27,000 K at 26 MHz with spectral index $-2.6$ and
corresponding dilution factor $W_{\nu}$ =1 (the drop of Galactic
background is expected to be at $\nu < 4$ MHz), and the term
determining contribution of Cas A emission to radiative line
broadening is estimated as $W_{\nu}T_{CasA}\approx$ 12,000 K at 26
MHz (which is comparable to the influence of Galactic background).

\begin{figure}
\includegraphics[width=84mm]{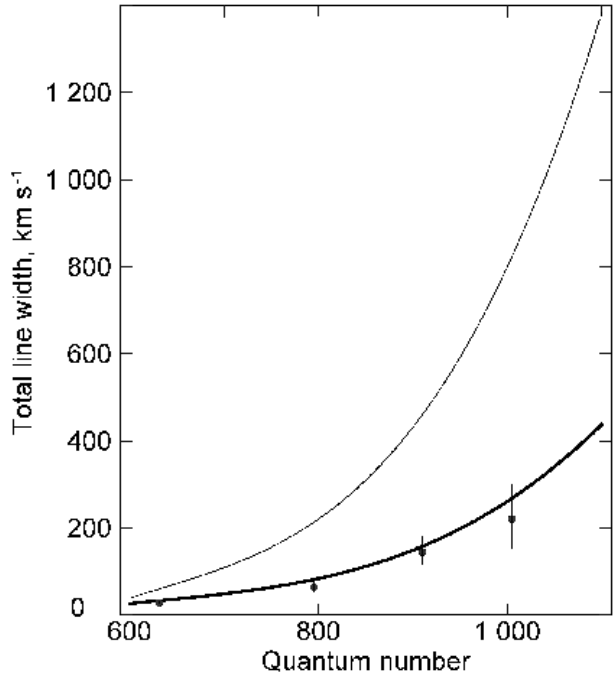}
\caption{The dependence of line width against quantum number. The
narrow line corresponds to the best previous model. The wide line
depicts our model which takes into account the drop of Cas A flux
density at $\nu < 20$ MHz.  Points depict experimental results.
Error bars were calculated using the noise statistics of the
data.}
\label{fig4}
\end{figure}

The pressure broadening \citep{b13} can be calculated using

\begin{equation}
 \Delta V_{p}=2*10^{-8}e^{-26/T^{1/3}_{e}}n_{e}n^{5.2}/T^{1.5}_{e}c/\nu, km s^{-1}.
\end{equation}

Correspondence between the curve predicted by our arguments and
measured parameters is good, whereas the older models overestimate
the line widths. When we take into account the typical conditions
of the ISM in the Galaxy, the fundamental limit of the highest
bound state of a Rydberg atom is near n=1,700. These are arrived
at by estimating the quantum level at which the rate of
depopulation exceeds the orbital frequency of the electron
\citep{b13,b3}. Our detection of the highest bound level at
n=1,005 (1,009) is thus well within this range and close to the
value quoted earlier \citep{b13}. This carbon atom is expected to
have a classical diameter of about 108 $\mu$ (~0.1 mm). Compare
this with the Bohr diameter ~ 0.000106 $\mu$. The carbon atom in
this tenuous nebula is larger by a factor of $\sim10^{6}$ compared
to the ground state atom. Taking into consideration the obtained
data and criterion that the distance between adjacent lines have
to be more than half the line width, we can expect that it will be
possible to observe the space atoms with n=1,100 - 1,200 in the
frequency range 15 - 20 MHz. Line blending is expected at larger n
because of the decreasing distance between the lines in
correspondence with the expression:

\begin{equation}
  \Delta \nu \approx(6R_{C}Z^{2}\Delta n)/n^{4} \approx 3\nu/n
\end{equation}
where $R_{C}$ is the Rydberg constant for carbon and Z is the
charge of an ion.

\section{Conclusions}
\label{sec4} In this paper, we have reported the detection of the
largest bound atom in space, which has recombined to quantum
levels of $\sim 1009$. The final spectrum  presented here shows
the presence of $\alpha$, $\beta$, $\gamma$ and $\delta$ lines.
This spectrum near 26 MHz was obtained by integrating data towards
Cas~A obtained using UTR-2 telescope in Ukraine.

At such low frequencies, the line widths are expected to increase
steeply ($\sim \propto n^5$) due to radiation and pressure
broadening which strongly dominate over thermal Doppler
broadening. However, we find that the observed increase is much
slower than predicted. We believe that this slow increase is due
to the decrease in the background radiation field at low
frequencies due to the spectral turnover in the non-thermal
spectra of Cas~A. After incorporating this in the model, we find
an excellent fit to the line widths of our observed data. Thus, we
suggest that this decrease in the field should be incorporated in
any model which explains the RRL data towards Cas~A.

Presently investigations
of low frequency RRLs with UTR-2 are continuing for many galactic
objects and data processing is progressing. It is clear that
RRLs studies will be important part of the scientific programs of
future giant low frequency instruments such as LOFAR.

\section*{Acknowledgments}

We acknowledge the efforts of late Prof. K.R. Anantharamaiah who
played a seminal role in initiating the joint Indo-Ukrainian
project which were funded by the Department of Science and
Technology, India and Ministry of Education and Science, Ukraine.
We thank Prof. W. Erickson for supplying the observatory UTR-2
with elements necessary to build the 4096-channel correlometer. We
thank Prof. D. P. Dewangan for bringing to our view the laboratory
work on Rydberg atoms. The works were carried out with partial
support of the grant INTAS 03-5727.

\label{lastpage}


\begin{thebibliography}{99}
\bibitem[\protect\citeauthoryear{Blake D.H., Crutcher R.M. \& Watson W.D.}
{1980}]{b1} Blake D.H., Crutcher R.M., Watson W.D., 1980, Nat.,
287, 707
\bibitem[\protect\citeauthoryear{Braude et al.}{1978}]{b2}
Braude S.Ya., Megn A.V., Ryabov B.P., Sharykin N.K., Zhouk I.N.,
1978, Astrophysics and Space Science, 54, 3
\bibitem[\protect\citeauthoryear{Brocklehurst M. \& Seaton M.J.}{1972}]{b3}
Brocklehurst M., Seaton M.J., 1972, MNRAS, 157, 179
\bibitem[\protect\citeauthoryear{Connerade, J.-P}{1998}]{b4}
Connerade, J.-P., 1998, Highly Excited Atoms. Cambridge Univ. Press, Cambridge
\bibitem[\protect\citeauthoryear{Gordon M.A. \& Sorochenko R.L.}{2002}]{b5}
Gordon M.A., Sorochenko R.L., 2002,Radio Recombination Lines:
Their Physics and Astronomical Application. Kluwer, Dordrecht
\bibitem[\protect\citeauthoryear{Kantharia N.G., Anantharamaiah K.R. \& Payne H.E.}
{1998}]{b6}Kantharia N.G., Anantharamaiah K.R., Payne H.E., 1998,
ApJ, 506,758
\bibitem[\protect\citeauthoryear{Konovalenko A.A. \&  Sodin L.G.}{1980}]{b7}
Konovalenko A.A., Sodin L.G., 1980, Nat., 283, 360
\bibitem[\protect\citeauthoryear{Konovalenko A.A. \&  Sodin L.G.}{1981}]{b8}
Konovalenko A.A., Sodin L.G., 1981, Nat., 294, 135
\bibitem[\protect\citeauthoryear{Konovalenko A.A.,
Stepkin S.V. \&  Shalunov D.V.}{2002}] {b9} Konovalenko A.A.,
Stepkin S.V., Shalunov D.V., 2002, in Pramesh Rao A.,  Swarup G.,
Gopal-Krishna eds, Proc. IAU Symp.199, The Universe at Low Radio
Frequencies. Astron. Soc. Pac., Pune, India, p. 349
\bibitem[\protect\citeauthoryear{Payne H.E., Anantharamaiah K.R. \& Erickson W.C.}{1989}]
{b10} Payne H.E., Anantharamaiah K.R., Erickson W.C., 1989, ApJ,
341, 689
\bibitem[\protect\citeauthoryear{Payne H.E., Anantharamaiah K.R. \& Erickson W.C.}{1994}]
{b11} Payne H.E., Anantharamaiah K.R., Erickson W.C., 1994, ApJ,
430, 690
\bibitem[\protect\citeauthoryear{Salem M. \& Brocklehurst M.}{1979}]{b12} Salem M.,
Brocklehurst M., 1979, ApJS, 39, 633
\bibitem[\protect\citeauthoryear{Shaver P.A.}{1975}]{b13}
Shaver P.A., 1975, Theoretical intensities of low frequency
recombination lines. Pramana 5, 1
\bibitem[\protect\citeauthoryear{Sorochenko R.L. \& Walmsley C.M.}{1991}]{b14}
Sorochenko R.L., Walmsley C.M., 1991, Astronomy Astrophysics
Transactions, 1, 31
\bibitem[\protect\citeauthoryear{Walmsley C.M. \& Watson W.D.}{1982}]{b15}
Walmsley C. M., Watson W. D., 1982, ApJ, 260, 317
\end{thebibliography}
\end{document}